\documentclass[iop]{emulateapj}
\usepackage{amsmath}

\begin{document}

\title{Accounting for Chromatic atmospheric effects on Barycentric Corrections}

\author{Ryan T. Blackman, Andrew E. Szymkowiak, Debra A. Fischer, Colby A. Jurgenson}
\affiliation{Department of Astronomy, Yale University, 52 Hillhouse Ave, New Haven, CT 06511, USA; ryan.blackman@yale.edu}


\begin{abstract}
Atmospheric effects on stellar radial velocity measurements for exoplanet discovery and characterization have not yet been fully investigated for extreme precision levels. 
We carry out calculations to determine the wavelength dependence of barycentric corrections across optical wavelengths, due to the ubiquitous variations in air mass during observations. 
We demonstrate that radial velocity errors of at least several cm s$^{-1}$ can be incurred if the wavelength dependence is not included in the photon-weighted barycentric corrections.
A minimum of four wavelength channels across optical spectra (380-680 nm) are required to account for this effect at the 10 cm s$^{-1}$ level, with polynomial fits of the barycentric corrections applied to cover all wavelengths.
Additional channels may be required in poor observing conditions or to avoid strong telluric absorption features. 
Furthermore, consistent flux sampling on the order of seconds throughout the observation is necessary to ensure that accurate photon weights are obtained. 
Finally, we describe how a multiple-channel exposure meter will be implemented in the EXtreme PREcision Spectrograph (EXPRES). 
\end{abstract}

\section{Introduction}
Thousands of exoplanets have been discovered in the past two decades through Doppler measurements and transit photometry. 
These techniques have been enabled primarily by the development of new technologies for spectroscopic and photometric observations. 
Radial velocity measurements are biased toward massive planets in short-period orbits, but improvements in precision have led to the discovery of super-Earths and Neptune-like planets. 
Transit photometry with the \textit{Kepler} and \textit{COnvection ROtation and planetary Transits} spacecrafts has enabled the detection of smaller exoplanets, and many more are expected with the next generation of space-based photometry missions such as the Transiting Exoplanet Survey Satellite \citep[TESS;][]{ricker2014}, the CHaracterizing ExOPlanet Satellite \citep[CHEOPS;][]{fortier2014}, and the PLAnetary Transitions and Oscillation of stars \citep[PLATO;][]{rauer2014}. 
The radial velocity and transit techniques complement each other by enabling the calculation of bulk planet densities, as mass can be determined from the radial velocity semi-amplitude and radius can be determined from the transit depth.
Currently, one of the key goals in exoplanet science is to detect Earth analogues and statistically assess their prevalence around main-sequence stars. 
Further advances in strategy, analysis, and instrumentation are necessary to begin the discovery of these exoplanets, and the radial velocity technique is a viable path for reducing the observational bias toward large planets in the current known sample \citep{mayor2014,fischer2016}.

It will not be possible to measure the masses for the majority of current and future transit detections or discover Earth analogues without significant improvement in radial velocity precision. 
Earth-like planets induce a reflex velocity on their stellar hosts with an amplitude on the order of 10 cm s$^{-1}$, requiring a factor of 10 in improvement in radial velocity precision over the current state of the art. 
The most significant terms in the error budget for radial velocity measurements come from photospheric velocities and instability in the instruments. 
The upcoming generation of spectrographs optimized for Doppler measurements aim to reach instrumental precisions that are sufficient to detect Earth analogues orbiting small stars \citep{pepe2010,jurgenson2016,schwab2016,sz2016}. 
Achieving this level of precision requires an extremely stable instrument, more precise wavelength calibration techniques, and techniques for modeling telluric contamination in the observed spectra.

The precise analysis of stellar velocities also requires correction for the motion of the observatory with respect to the solar system barycenter. 
This correction is applied to the measured Doppler shift multiplicatively through
\begin{equation}
z_{\mathrm{true}} = (1+z_{\mathrm{meas}})(1+z_{B})-1, 
\end{equation} 
where $z_{\mathrm{true}}$ is the Doppler shift of the host star relative to the barycenter of the system, $z_{\mathrm{meas}}$ is the measured Doppler shift, and $z_{B}$ is the barycentric correction \citep{we2014}. 
These Doppler shifts are related to the change in photon wavelength through
\begin{equation}
\lambda_\mathrm{obs} = (1+z)\lambda_\mathrm{ref},
\end{equation}
where $\lambda_{\mathrm{ref}}$ is the reference wavelength and $\lambda_{\mathrm{obs}}$ is the observed, shifted wavelength.
The systems engineering error budgets for the next generation of Doppler spectrographs typically allocate barycentric correction errors of no more than 2 cm s$^{-1}$ \citep{podgorski2014,jurgenson2016,halverson2016}. 
One complication is that the barycentric correction can only be calculated for an instant in time, but exposures have durations up to about 30 minutes.  
Because the barycentric correction does not strictly change linearly in time, corrections should be calculated throughout an observation and applied as a flux-weighted average.
For a ground-based observation, atmospheric extinction will introduce a wavelength dependence in the transmittance of photons to the instrument, and therefore, a wavelength dependence in the barycentric correction. 

Here, we carry out simulated observations to investigate the importance of including this chromatic dependence, due to air mass variations, in the barycentric correction in preparation for the commissioning of the EXtreme PREcision Spectrograph (EXPRES) in 2017 \citep{jurgenson2016}. We discuss how transient effects such as variations in the optical depth of different atmospheric components may compound this effect.
In addition, we discuss important considerations for the instrumentation requirements of the multiple-channel exposure meters that will be used to correct for this effect. 

\section{Simulated Observations and Barycentric Corrections}\label{sec:sim}
\subsection{Atmospheric Transmittance}
\begin{figure}[]\label{fig:am}
\centering
\includegraphics[scale=1.0]{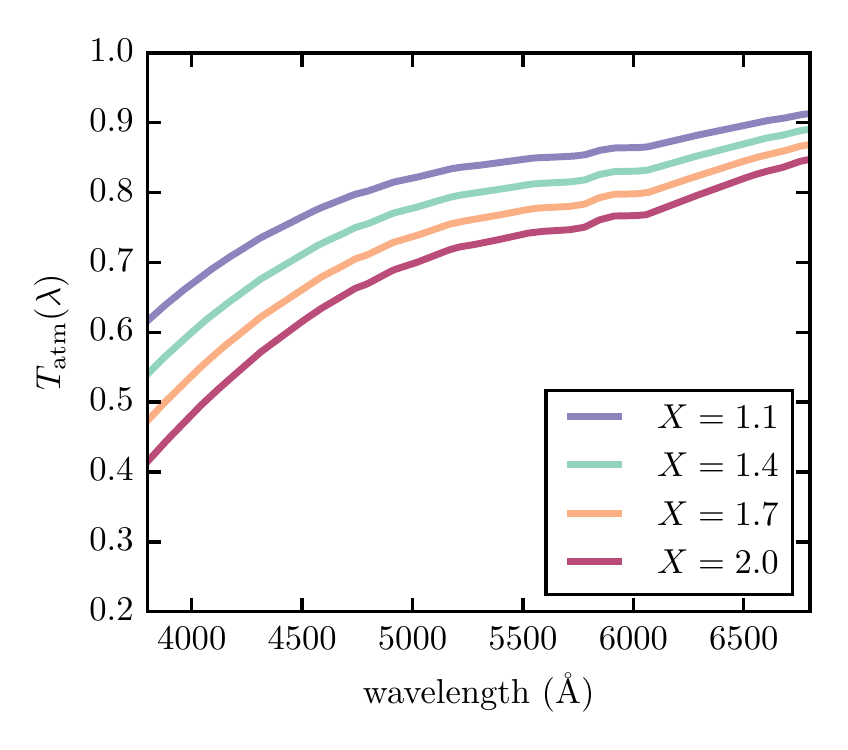}
\caption{Mean atmospheric extinction law for Kitt Peak National Observatory for different air mass values.}
\label{am}
\end{figure}

\begin{figure}[]\label{fig:atmos}
\centering
\includegraphics[scale=1.0]{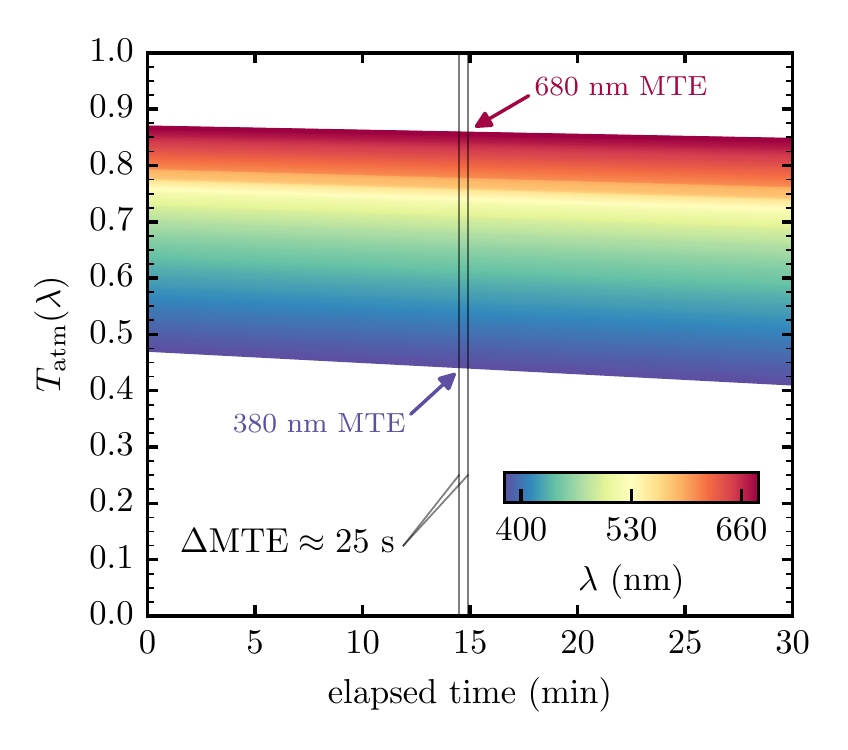}
\caption{Example of how the atmospheric transmittance changes throughout a 30 minute observation of a setting object at an initial air mass of 1.7 and a final air mass of 2.0, in optical wavelengths. The fractional change in transmittance is larger for bluer wavelengths. The mean time of the exposure (MTE) for blue and red wavelengths is separated by about 25 s.}
\label{atmos}
\end{figure}

We simulate the effect of an atmosphere on the calculations of barycentric corrections by using the mean extinction law derived for Kitt Peak National Observatory (KPNO) provided with the Image Reduction and Analysis Facility (IRAF)\footnote{IRAF is distributed by the National Optical Astronomy Observatories, which are operated by the Association of Universities for Research in Astronomy, Inc., under cooperative agreement with the National Science Foundation.} software package \citep{tody93}. 
Typical exposure times for radial velocity measurements of exoplanet host stars are tens of minutes long, which are the timescales that are most relevant for our simulations. 
Wavelength-dependent changes in the transmittance of the atmosphere determine how the observed stellar flux will change throughout an observation. 
The transmittance of the atmosphere is dominated by Rayleigh scattering in blue wavelengths, but contributions from aerosol content and ozone are significant at redder wavelengths \citep{buton2013A}.
With knowledge of how the air mass changes throughout an observation, it is possible to calculate how the transmittance changes throughout that observation with
\begin{equation}
T_{\mathrm{atm}}(\lambda) = e^{-X\kappa(\lambda)},
\end{equation}
where $T_{\mathrm{atm}}(\lambda)$ is the transmittance of the atmosphere, $X$ is the air mass, and $\kappa(\lambda)$ is the extinction coefficient of the atmosphere per air mass. When determining the air mass throughout the simulated observations, we adopt the analytical form introduced by \citet{young1967}  as
\begin{equation}
X \approx \sec (z) [1-0.0012(\sec^2 (z) -1)],
\end{equation}
where $z$ is the zenith angle. This expression is accurate to $80\degr$ zenith angles. 
Atmospheric transmittance curves at different air mass values for the KPNO mean extinction law are shown in Figure~\ref{am}.
An example of changing atmospheric transmittance is shown in Figure~\ref{atmos} for a 30 minute observation of a setting object in the due west direction at an initial air mass of 1.7 and a final air mass of 2.0. 
Observations reaching this air mass are generally discouraged, but must occasionally be performed in radial velocity surveys in order to obtain observations of target stars at high cadence.
Therefore, this simulated observation represents a situation close to a worst-case scenario in terms of sky position.
This example illustrates that the fractional change in transmittance throughout an observation is highly dependent on wavelength, as blue light is preferentially attenuated as air mass increases. 
The photon-weighted mean time of the exposure (MTE) of the observation at 380 nm occurs approximately 25 s earlier than at 680 nm.
A time difference this large will correspond to a difference in barycentric velocity of tens of cm s$^{-1}$ \citep{we2014}.

\subsection{Barycentric Corrections}
\begin{figure}[]\label{fig:spectra}
\centering
\includegraphics[scale=1.0]{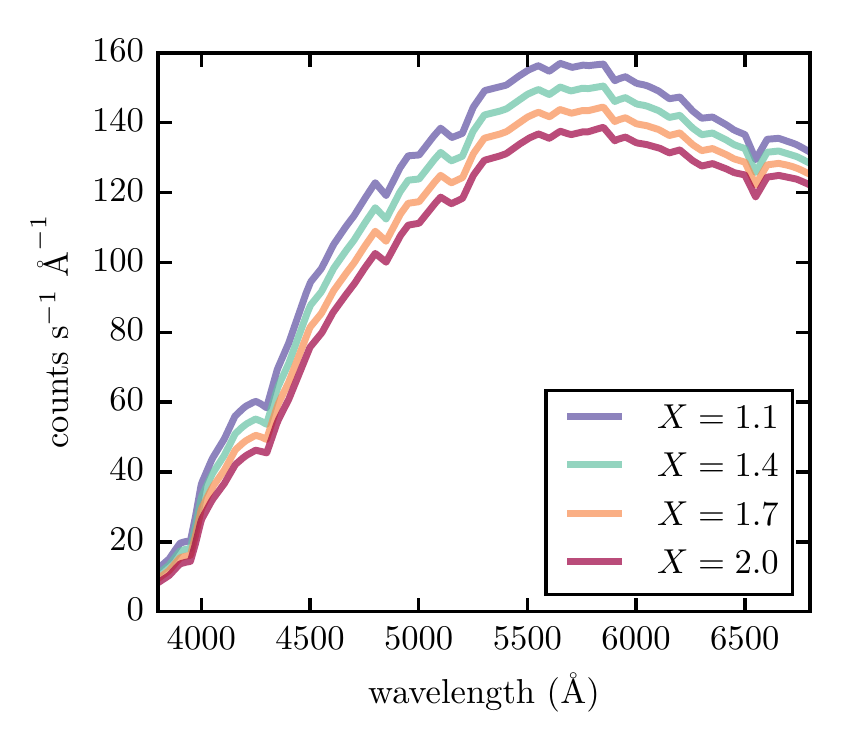}
\caption{Spectrum of the G0 star LTT 7379, as it would be detected by the exposure meter of EXPRES at different air mass values.}
\label{spectra}
\end{figure}

In our simulations, we propagate the spectrum of a star through the atmosphere, telescope, and instrument, and then calculate the barycentric correction weights from the signal received at the exposure meter of the instrument. 
The stellar spectrum we adopt is that of LTT 7379, a G0 main-sequence star \citep{hamuy1992}.
The simulated observations are carried out at the location of Lowell Observatory in Happy Jack, AZ, USA.
We also adopt the instrumental throughput determined for the Discovery Channel Telescope and the predicted throughput to the exposure meter of EXPRES, including the quantum efficiency of the detector \citep{jurgenson2016}. 
Instrumental losses should not change over time, but these assumptions yield a more realistic simulation and should not affect the generalization of the results to other spectrographs. 
The G0 stellar spectrum of LTT 7379 is shown at different air masses in Figure \ref{spectra} as it would be detected by the exposure meter of EXPRES.
Signal weights are calculated for the effective wavelength midpoint of each channel of the exposure meter.
Because the spectrum of every star is different, the effective wavelength midpoint of each channel in the exposure meter will be different for every observation, and is found with
\begin{equation}
\lambda_{\textrm{eff}} = \frac{\int \lambda F(\lambda) \textrm{d}\lambda}{\int F(\lambda) \textrm{d}\lambda},
\end{equation}
where $F(\lambda)$ is the spectrum of light that is detected and the integrals are taken over the wavelength bounds of each channel. 
This spectrum depends on the properties of the star being observed, but is also dependent on the throughput of all optical components of the telescope and instrument because the transmission and reflection efficiencies of every component are wavelength dependent. 

We calculate the barycentric corrections for simulated observations with the publicly available code \texttt{BARYCORR}, which has been tested to be accurate at the 1 cm s$^{-1}$ level before complications arise from long exposure times and atmospheric effects \citep{we2014}. 
The required inputs for reaching this precision with \texttt{BARYCORR} are the measured Doppler shift, exposure time as a Julian date, coordinates of the star, location and elevation of the observatory, proper motion, parallax, and bulk radial velocity of the star system. 
In our simulations, we calculate a barycentric correction for each time sampled with the exposure meter as discussed in \citet{fischer2016}.
These barycentric corrections are then weighted by the number of photons detected in each channel to obtain the final barycentric correction at each effective wavelength midpoint. 
The number of photons used for the weights for every time sample $i$ and each channel $k$ is found with
\begin{equation}
w_{k,i} = \int_{t_{0,i}}^{t_{f,i}} \int_{\lambda_{0,k}}^{\lambda_{f,k}} F(\lambda) T_{\mathrm{atm}}(\lambda) A_{\mathrm{tele}} T_{\mathrm{inst}}(\lambda) \frac{\lambda}{hc} \mathrm{d}\lambda \mathrm{d}t,
\end{equation}
where $A_{\mathrm{tele}}$ is the effective area of the primary mirror of the telescope, $T_{\mathrm{inst}}(\lambda)$ is the efficiency of all optical components leading up to the detector and including the quantum efficiency of the detector, $\lambda_0$ and $\lambda_f$ are the wavelength bounds on each channel, and $t_0$ and $t_f$ are the initial and final times over which the exposure meter integration takes place.
The weighted barycentric correction for each channel is then calculated with 
\begin{equation}
z_{B,k} = \frac{\sum_{i=1}^{n}(z_{B,i}w_{k,i})}{\sum_{i=1}^{n}(w_{k,i})},
\end{equation}
where $n$ is the total number of exposure meter integrations.
The barycentric corrections at the effective wavelength midpoints of each channel can then be fit by a polynomial to obtain corrections at every wavelength across a spectrum.
This method differs from the previously used method in which the weighted MTE is used for calculating barycentric corrections \citep[see][]{landoni2014, we2014}. 
\citet{fischer2016} and references therein discuss that taking weighted barycentric corrections may improve corrected radial velocities by up to 25 cm s$^{-1}$ compared to the method of using the weighted MTE.

\section{Results}
\begin{figure*}
\centering
\includegraphics[scale=1.0]{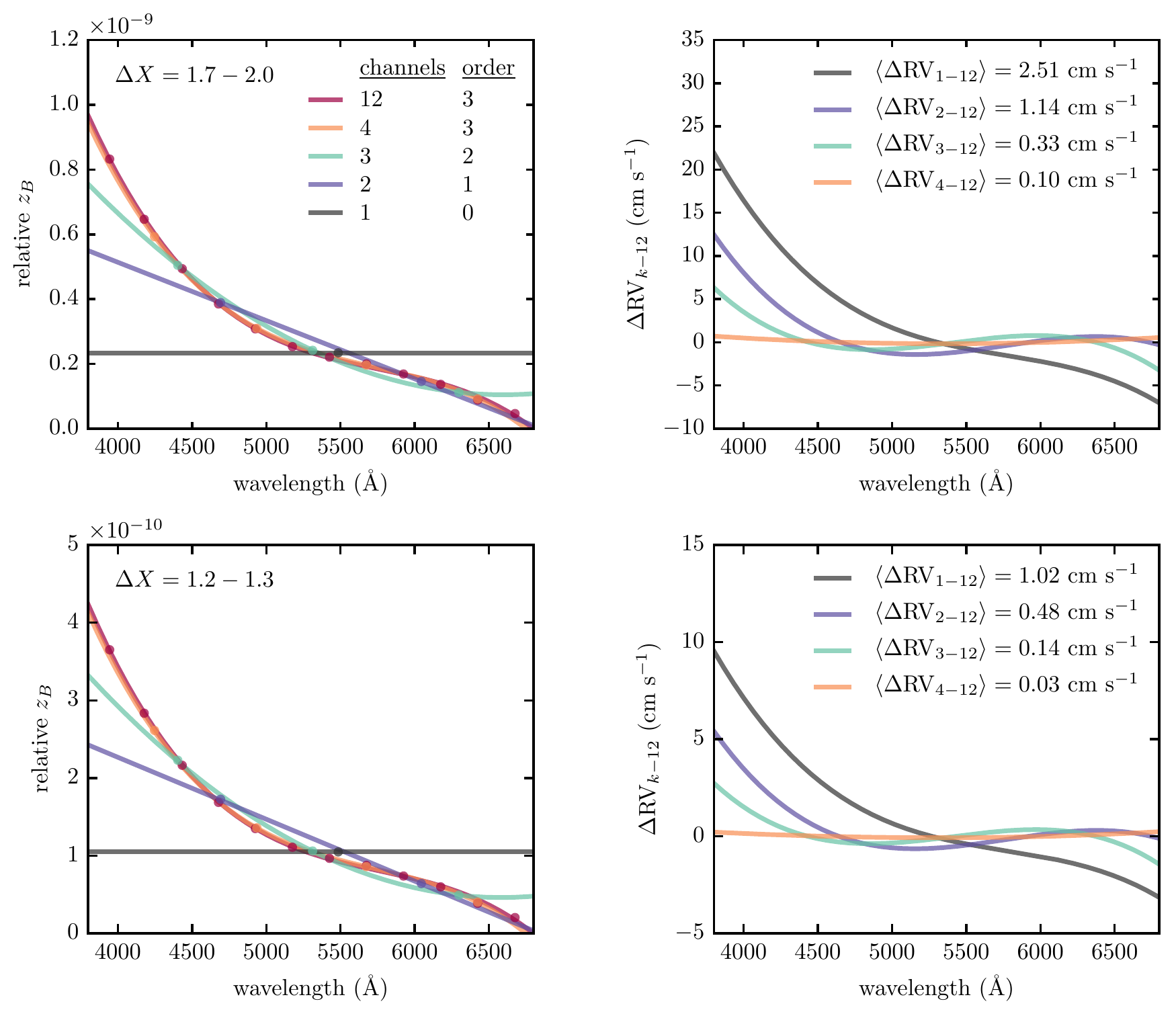}
\caption{Left: relative barycentric corrections as a function of wavelength for two simulated 30 minute exposures of a spectral type G0 star at an initial air mass of 1.7 and final air mass of 2.0 (top) and an initial air mass of 1.2 and final air mass of 1.3 (bottom). The number of channels and the polynomial-fit order for each simulation is shown in the legend. Right: radial velocity (RV) values relative to the twelve-channel results for the barycentric correction values shown in the left panels. The legend indicates the average RV offset compared to the twelve-channel case.}
\label{zbs_rvs} 
\end{figure*}

We determine the optimal number of channels for the exposure meter by calculating the signal weights for simulated observations with different numbers of channels. 
The results are shown in Figure \ref{zbs_rvs} for two observations, one from an air mass of 1.7-2.0 (top two panels) and one from an air mass of 1.2-1.3 (bottom two panels).
With a single channel exposure meter, the barycentric correction is assumed to be the same for all wavelengths.  
For an exposure meter with two channels, a linear fit can be applied to two barycentric corrections for interpolation across the spectrum. 
With additional channels, higher-order polynomial fits can be applied to improve the accuracy of interpolated barycentric corrections. 
The red curves in the left panels of Figure \ref{zbs_rvs} show relative barycentric corrections for 12 channels and a third-order polynomial fit, which displays a behavior in the corrections that is not accounted for by lower order fits. 
The individual channel barycentric corrections are well-fit by this polynomial, indicating that a third-order fit is sufficient to account for the chromatic dependence. 
The orange curves in Figure \ref{zbs_rvs} show a third-order fit with only four channels, and is nearly identical to the twelve-channel result.
The two right panels show the corresponding velocity differences that are incurred as a function of wavelength relative to the oversampled case with 12 channels, with the average offset labeled in the legends.
The difference in calculated radial velocity on the blue end compared to the red end of the spectrum reaches nearly 30 cm s$^{-1}$ for the higher air mass case. 
This indicates the importance of using multiple channels to account for the wavelength dependence of the barycentric correction. 
The radial velocity error for an observation incurred by not including the chromatic dependence in the barycentric correction will depend on which lines are used for the Doppler analysis and where they reside in the spectrum.
In the high air mass example, an even distribution of lines across the spectrum taken with equal weights will result in a radial velocity error of 2.5 cm s$^{-1}$ for a single channel exposure meter compared to the twelve-channel case. However, the lines used in the Doppler analysis are generally not evenly distributed. 
For G and K dwarf stars, there are more lines in blue wavelengths than in the red, so the velocity measurement may be biased toward this side of the spectrum where the error is larger.

\begin{figure*}[]
\centering
\includegraphics[scale=1.0]{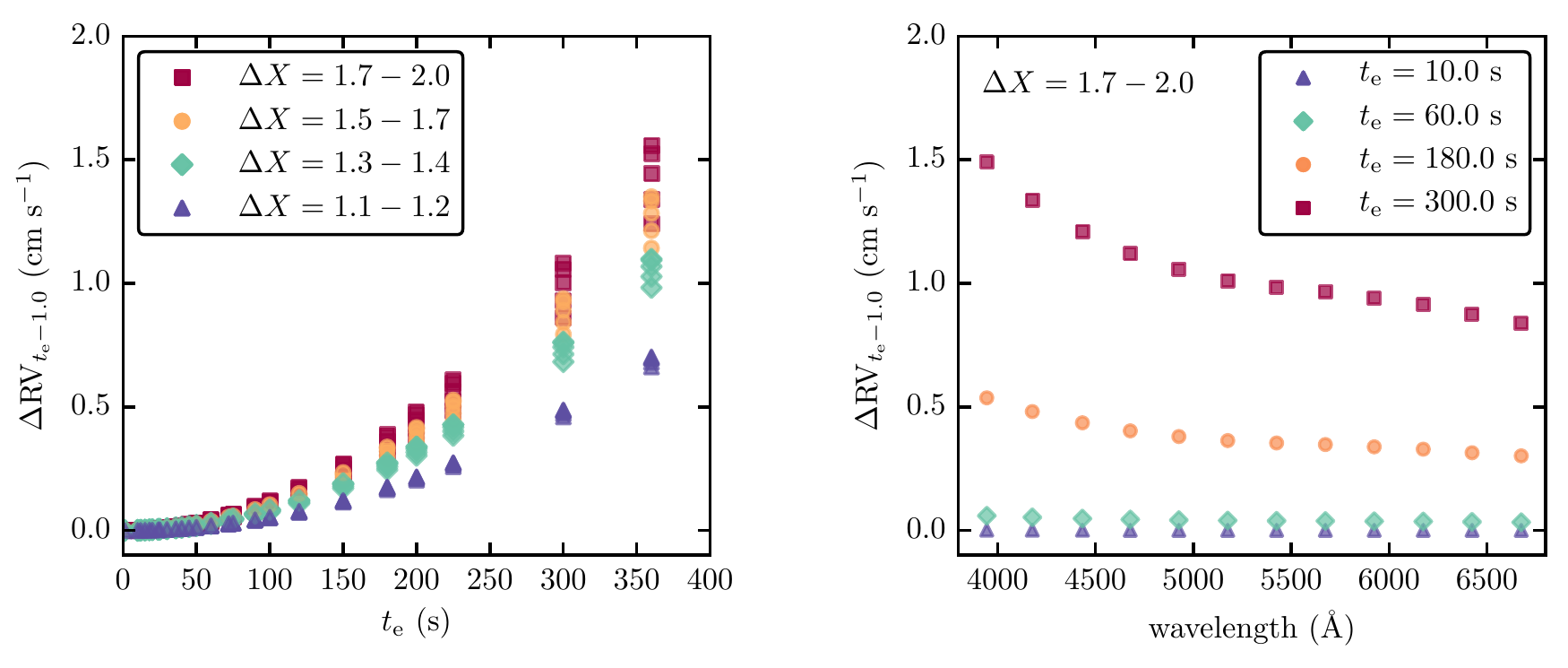}
\caption{Left: differences in radial velocity obtained from different exposure meter integration lengths ($t_{\mathrm{e}}$), relative to the 1.0 s integration length result. The observations are 30 minutes long at initial air masses ranging between 1.1 and 1.7 throughout a 10 hr night. Longer integration times of the exposure meter lead to less accurate barycentric corrections and significant errors in radial velocity. Right: radial velocities from different exposure meter integration lengths relative to the 1.0 s exposure meter integration length result for a single 30 minute observation at an initial air mass of 1.7 and final air mass of 2.0.}
\label{em_integration_errors}
\end{figure*}

We also quantify the radial velocity error induced by the length of the integration time chosen for the exposure meter. 
This error is incurred because the time stamp of every exposure meter integration is taken to be the midpoint. 
However, the midpoint of this integration will never be equivalent to the weighted MTE, due to the atmospheric effects described in this paper. 
Although this introduces an inherent error associated with this method, with short integration times of the exposure meter, the midpoint of the integration approaches the weighted MTE. 
This error is shown in the left panel of Figure \ref{em_integration_errors} for different exposure meter integration lengths at different air masses.
The simulated observations are 30 minutes long, in the due west direction, with initial air masses of each observation ranging from 1.1 to 1.7.
These simulated observations are then repeated every two hours over a 10 hr period representing a full night. 
The error is taken to be the average radial velocity offset between the 1.0 s integration result and the other integration lengths.
Appreciable radial velocity errors start to occur when the exposure meter integration length becomes longer than about two minutes.
The radial velocity errors obtained from a single observation in different exposure meter channels with different integration lengths are shown in the right panel of Figure \ref{em_integration_errors}.
The exposure meter integration length error is moderately dependent on wavelength, as larger errors are incurred in blue wavelengths. 

For every exposure meter integration throughout an observation, there will be some down time for the exposure meter to read out. 
This will introduce a small error in the signal measurements if left unaccounted for, but should be negligible given the extremely fast readout times of the modern detectors that will be used in future exposure meters.

\section{Discussion}\label{sec:discussion}
\subsection{Atmospheric Effects}
Atmospheric extinction induces a ubiquitous wavelength dependence in the barycentric corrections for ground-based observations.
The extent of this effect depends on the parameters of the observation.
The wavelength dependence is stronger when the observation passes through bigger differences in air mass. 
Therefore, observations performed near the meridian or at low air masses are less sensitive to this effect. 
Observations performed closer to Earth's equator will be more sensitive to this effect, because stars will traverse bigger changes in air mass compared to observations of the same duration made closer to Earth's poles. 
Longer observations, and ones performed in the easterly and westerly directions that will necessarily pass through larger differences in air mass, will be the most sensitive to this effect. 
The time of night and the time of year did not have a significant impact on the results. 
We find only slight differences in the results between G and M type stellar spectra; the general conclusions remain the same. 

Although the changing air mass during an observation induces a somewhat predictable change in atmospheric transmittance, there are also other potential transient wavelength-dependent effects on stellar flux including variable seeing, guiding errors, and variations in the optical depth of the different components of atmospheric attenuation (atomic and molecular scattering and absorption, aerosols, and clouds). 
For example, \citet{landoni2014} carried out simulations that showed how wavelength-dependent atmospheric perturbations can impact the accuracy of barycentric corrections. 
However, they did not consider the impact of changing atmospheric extinction throughout an observation, due to changing air mass, which is dependent on sky position and can introduce significant wavelength-dependent barycentric correction errors, even in stable observing conditions. 
Transient effects are more difficult to predict in advance of an observation, and their severity depends on the observing conditions and the specific instrumentation in use. 
The results we provide only take into account how the air mass changes throughout an observation, but otherwise we implicitly assume stable observing conditions. 
It is therefore possible that the wavelength dependence is more significant in more realistic scenarios. 
\citet{stubbs2007} discuss the expected variations of the components of atmospheric attenuation over different timescales in the context of precision photometry, but their conclusions can be generalized for insight regarding high-precision spectroscopy as well.

The extinction coefficient of atomic and molecular Rayleigh scattering is given by
\begin{equation}
\kappa_R(\lambda,P,h) = \frac{2.5}{\ln(10)} \frac{\sigma(\lambda) P}{g(h) M},
\end{equation}
where $P$ is atmospheric pressure, $M$ is the molecular mass of dry air, $g(h)$ is the acceleration due to gravity at altitude $h$, and $\sigma(\lambda)$ is the Rayleigh scattering cross section \citep{buton2013A}. 
\citet{rayleigh1871} found that the intensity of scattered light is proportional to $\lambda^{-4}$, indicating that this component is much more significant in blue wavelengths. 
However, changes in pressure occur on timescales of days \citep{stubbs2007}, and so will be mostly irrelevant for precision spectroscopic observations, which only last tens of minutes.

Molecules present in the atmosphere such as O$_2$, ozone, and water produce many absorption features in optical and near-infrared spectra. 
Any variations of these components during observations will have an impact on the chromatic transmittance of the atmosphere. 
Oxygen molecules produce strong absorption bands in red wavelengths, but oxygen content is expected to be axisymmetric about the zenith and stable over time \citep{stubbs2007}. 
Ozone impacts atmospheric transmittance at a level up to a few percent in the Chappuis Band \citep{chappuis1880} between wavelengths of 500 nm $< \lambda <$ 700 nm. 
Space-based instruments such as the \textit{Total Ozone Mapping Spectrometer} (TOMS) have measured ozone variability. 
\citet{stubbs2007} report annual cyclic variability of $\pm 25\%$ above the PanSTARRS observing site with much smaller nightly variations. 
Even extreme changes in ozone content would have only a small impact on chromatic exposure timing based on the figures presented in \citet{stubbs2007}. 
Water molecules produce significant absorption features in red and near-infrared wavelengths. 
Water content variations occur on both nightly and longer timescales \citep{thomas-osip2007,frogel1998} and the attenuation likely has an azimuthal dependence as well \citep{stubbs2007}, and thus could be an important factor in chromatic exposure timing given that stars are tracked across the sky during observations.
Near-infrared spectrographs will especially suffer from water absorption in the 800 nm to 1 $\mu$m range, and may need to avoid these features entirely in the Doppler analysis.  

Transient changes in aerosol content could also lead to chromatic changes in atmospheric transmittance on observation timescales. 
Long-term changes with moderate wavelength dependence in the Mie scattering regime have clearly been observed, and have been correlated with seasons as well as surface events on Earth such as forest fires and volcanic eruptions \citep{holben2001,burki1995}. 
These studies also present evidence for daily variations, indicating a potential impact on atmospheric transmittance on timescales of observations. 
\citet{stubbs2007} note that wind direction can impact the effect of aerosol attenuation, adding an azimuthal component to transmittance. 
However, \citep{buton2013A} found no significant evidence for aerosol content variations on nightly timescales at Maunakea.

Clouds composed of water droplets and ice crystals can significantly reduce atmospheric transmittance on timescales relevant to astronomical observations. 
However, these particles are generally larger than the wavelengths of optical light, and so scattering should be independent of wavelength \citep{buton2013A}. 
Non-chromatic atmospheric effects such as clouds are not relevant for chromatic measurements of atmospheric attenuation, but the exposure meter integration length must be set to account for the variation in attenuation across all wavelengths during the observations. 
Given that clouds may pass through an observation on timescales of minutes, integration times of the exposure meter on the order of seconds should be able to completely account for this effect.

It is possible, if not probable, that using more than four channels will be necessary to accurately account for the wavelength dependence of transient atmospheric effects on barycentric corrections. The use of 12 channels would provide a more sound approach in substandard observing conditions, as the spectrum would be more frequently sampled. Higher-order polynomial fits may also be required. In addition to more channels, more rapid sampling with the exposure meter may also be required in substandard conditions. Given that the variations in the optical depth of the various atmospheric components can occur on timescales of minutes, integration lengths on the order of seconds may be a more appropriate sampling frequency. One limiting factor in the speed of sampling is how much signal can be detected in the exposure meter integration, as a sufficient number of photons is needed to exceed read noise.

\subsection{Exposure Meter of EXPRES}
We have established specific requirements for accurate chromatic exposure meters. 
The exposure meter of EXPRES will consist of an $R\approx250$ Czerny-Turner spectrograph and an electron-multiplying CCD (EMCCD) for the detector. 
An EMCCD multiplies the signal with a solid state device before readout so that readout noise remains small compared to the signal even in low-light conditions (see \citealt{daigle2014} and references therein). 
High signal-to-noise ratio in the exposure meter spectra will help ensure that accurate signal weights are obtained for the barycentric corrections. 
These devices can read out at speeds up to hundreds of frames s$^{-1}$ depending on the binning mode, which will be sufficient for any desired sampling frequency. 
EMCCD detectors are already commonly used for other high sampling frequency imaging applications in astronomy, such as with fast tip-tilt and wavefront-sensing adaptive optics systems \citep[e.g.,][]{young2014}. 

The exposure meter pickoff in EXPRES is behind the slit, and diverts 2\% of light to the exposure meter. 
It must be ensured that the exposure meter faithfully samples the same spectrum that is detected at the main CCD. 
This will be accomplished by measuring the chromatic sensitivity of the different optical paths and deriving correction factors that will be applied to the exposure meter spectra. 
We do not expect temporal variations between the exposure meter and the main spectrograph, due to high stability in pressure and temperature that the EXPRES vacuum enclosure provides. 

One important advantage of a spectrograph and CCD-based exposure meter is that the detected spectra can be binned into the desired number of channels for calculating weighted barycentric corrections. 
These channels need not be continuous, and can be chosen to avoid strong telluric absorption lines. 
These features are generally avoided in the Doppler analysis, and so they should be avoided in the exposure meter data to preserve appropriate weights for the barycentric corrections. 
The exposure meter spectrograph of EXPRES will be wavelength-calibrated with a ThAr emission lamp to ensure that wavelength channels are established with sufficient accuracy. 

\section{Conclusion}\label{sec:conclusion}
Future searches for small, rocky planets will rely on Doppler measurements of stellar hosts, which will complement upcoming space-based missions for transit photometry. 
Improvement in radial velocity precision will allow for the detection and characterization of these smaller planets. 
Error budgets for the next generation of instruments are very stringent, and currently assume the error from the barycentric correction is not larger than 2 cm s$^{-1}$. 
The algorithms for achieving this accuracy are complete, but no instrument has yet implemented the necessary components to reach the required level of precision. 

We have demonstrated the importance of calculating barycentric corrections as a function of wavelength due to variable atmospheric extinction throughout an observation.
The next generation of Doppler spectrographs that aim for the highest possible precision will need to include wavelength sampling by use of multiple-channel exposure meters.  
Wavelength sampling with at least four broadband channels and third-order polynomial fitting is sufficient to reduce chromatic errors in the barycentric correction to less than 1 cm s$^{-1}$. 
However, additional channels would provide a more sound approach when observing conditions are substandard. 
An exposure meter detector that reads out on the order of seconds will ensure that the error induced from the barycentric correction remains under 1 cm s$^{-1}$, but in stable atmospheric conditions, this interval could be extended. 

\acknowledgments
We thank the anonymous referee for helpful comments that improved this paper.
Support for this work was provided by the National Science Foundation under grant NSF MRI 1429365.
We acknowledge an insightful presentation by Jason Eastman and Lars Buchhave at the Extreme Precision Radial Velocities II conference which helped inspire this work. 
We also thank Jason Wright and Sagi Ben-Ami for helpful discussion.
This research made use of Astropy, a community-developed core Python package for Astronomy \citep{astropy2013}.

\bibliographystyle{aasjournal.bst}

\end{document}